\documentclass[11pt]{article}
\usepackage[T1]{fontenc}
\usepackage[utf8]{inputenc}
\usepackage{lmodern}
\usepackage[a4paper,margin=2.5cm]{geometry}
\usepackage{amsmath}
\usepackage{graphicx}
\usepackage{booktabs}
\usepackage[numbers,sort&compress]{natbib}
\usepackage[hidelinks]{hyperref}
\usepackage{xurl}

\title{Early Detection of Pension-Sponsor Distress from Public Filings:\\
A Sponsor Vulnerability Index, Stress-Augmented VaR,\\ and Dynamic Self-Investment Caps}
\author{Ravil Akhtyamov\\
\small Digital Economy Lab\\
\small \href{mailto:latam@digitaleconomylab.org}{latam@digitaleconomylab.org} \quad
ORCID \href{https://orcid.org/0000-0002-0783-8573}{0000-0002-0783-8573}}
\date{}

\begin{document}
\maketitle

\begin{abstract}
A defined-benefit pension is only as safe as its sponsor, yet sponsor distress is under-instrumented relative to plan-side modelling. We present a transparent, reproducible early-warning pipeline built entirely on public data. A \emph{Sponsor Vulnerability Index} (SVI)---a gradient-boosted classifier over point-in-time SEC EDGAR balance-sheet fundamentals and two macro series---predicts a filings-based distress state one to two quarters ahead. On 36 U.S. sponsors in five sector groups (1{,}418 firm-quarters, 2015\,Q1--2025\,Q3) it reaches an out-of-fold AUC-ROC of $0.887$ and PR-AUC of $0.711$ (base rate $11.9\%$) under leave-one-firm-out cross-validation. A Monte Carlo VaR with a stylized transition-stress overlay shows that climate-type stress is negligible for a diversified plan but adds $20$--$43\%$ to one-year $95\%$ VaR when $22.5\%$ of assets are held in sponsor securities. A dynamic self-investment cap maps the SVI to a limit that falls from $8\%$ to below $1\%$ in acute distress and recovers with the sponsor, which we illustrate on PG\&E's 2019 bankruptcy and Boeing's persistent fragility. We report the method's failure modes alongside its results---book-equity labels that also fire on non-credit events such as spin-offs, positives concentrated in one firm, and stylized stress parameters---and release code, data and results under a concept DOI.
\end{abstract}

\noindent\textbf{Keywords:} pension solvency; sponsor covenant; financial distress prediction; gradient boosting; Value-at-Risk; climate transition risk; employer securities.

\section{Introduction}
A defined-benefit (DB) pension promise is only as sound as the sponsor that backs it. When a corporate sponsor weakens, the plan inherits that weakness through three correlated channels: rising underfunding, concentrated holdings of the sponsor's own securities (\emph{self-investment}), and the joint deterioration of sponsor and plan in the same state of the world. Sponsor and plan risk are linked in both directions: plan risk is reflected in sponsor equity risk~\cite{jin2006}, and sponsor financial condition shapes funding and investment policy~\cite{sharpe1976,rauh2006,rauh2009}. Plan-side asset--liability modelling is mature~\cite{martellini2012}, but the operational question on the sponsor side---\emph{can distress be detected early enough for a trustee or supervisor to act?}---is comparatively under-instrumented. In practice covenant strength often enters as a slowly reviewed qualitative grade, and self-investment is governed by static caps such as the 10\% limit on employer securities in U.S.\ DB plans~\cite{erisa407} or the 5\% employer-related investment limit in the U.K.~\cite{ukeri}.

This paper presents a public-data early-warning pipeline that addresses this gap in three parts. A \textbf{Sponsor Vulnerability Index (SVI)} predicts, one to two quarters ahead, whether a sponsor will enter a distress state defined from its own filings. An \textbf{Enhanced VaR} engine prices the interaction of self-investment concentration with a transition-risk stress overlay. A \textbf{dynamic self-investment cap} converts the SVI into a limit that tightens as the sponsor weakens and relaxes as it recovers.

\paragraph{Contributions.}
(1)~A look-ahead-safe SVI trained on SEC EDGAR XBRL fundamentals for 36 U.S.\ sponsors, evaluated with leave-one-firm-out cross-validation so that every reported metric refers to firms unseen in training. (2)~A Monte Carlo analysis showing that transition-type stress and self-investment \emph{interact}: the stress add-on is close to zero without self-investment and grows with concentration. (3)~A transparent cap rule linking the SVI to a regulatory-style limit, illustrated on real episodes. (4)~A fully public pipeline (code, extracted filings, results) released for replication, together with an explicit account of where the method fails.

\section{Related work}
\paragraph{Distress prediction.} Accounting-ratio models of corporate failure run from univariate ratios~\cite{beaver1966} and discriminant analysis~\cite{altman1968} through logit~\cite{ohlson1980} and hazard models with market inputs~\cite{shumway2001,campbell2008}. Machine-learning methods, gradient-boosted trees in particular, are competitive on this task~\cite{barboza2017,jones2017}, consistent with broader evidence that tree ensembles remain strong on tabular data~\cite{grinsztajn2022,shwartzziv2022}. We use XGBoost~\cite{chen2016} with gain-based feature attribution; SHAP values~\cite{lundberg2017} are a natural extension. Our target differs from bankruptcy: we predict a book-equity distress state relevant to pension covenant, using only balance-sheet data that every listed sponsor files.

\paragraph{Sponsor risk and pensions.} Corporate finance treats the plan as part of the sponsor's balance sheet~\cite{sharpe1976,jin2006}; financially constrained or weak sponsors fund and invest differently~\cite{rauh2006,rauh2009}. Concentrated holdings of employer stock are a well-documented hazard in defined-contribution plans~\cite{benartzi2001}; in DB plans they are capped by statute~\cite{erisa407,ukeri}. Liability-driven investing typically treats sponsor strength as a static input~\cite{martellini2012}. We make it a time-varying, data-driven signal.

\paragraph{Climate and supervisory stress testing.} The NGFS scenario family supplies transition pathways for supervisory stress tests~\cite{ngfs2023}; climate stress tests of the financial system~\cite{battiston2017} and evidence that carbon risk is priced~\cite{bolton2021} motivate including transition stress in portfolio risk. Pension supervisors have begun to run such exercises, e.g.\ EIOPA's 2019 IORP stress test with an ESG component~\cite{eiopa2019} and the Bank of Russia's 2026 stress-testing scenarios for non-state pension funds~\cite{cbr2026}. Our overlay is deliberately simpler than these (Section~\ref{sec:var}); its role is to show the interaction with self-investment, not to forecast climate losses.

\paragraph{Prior version.} A Russian-language article by the author~\cite{akhtyamov2026} introduced the framework on a five-sponsor pilot (AUC $0.872$). The present paper re-estimates it on a 36-sponsor panel under leave-one-firm-out validation and supersedes the pilot numbers.

\section{Data}\label{sec:data}
\subsection{Sponsor fundamentals}
We query the SEC EDGAR XBRL \texttt{companyconcept} API~\cite{edgar} for quarterly instantaneous frames of \texttt{Assets} and \texttt{StockholdersEquity} for 41 U.S.\ sponsors, and keep firms with at least 16 quarters in which both are reported. Five firms (CAT, T, VZ, JNJ, PG) drop out because they tag equity only under the concept that includes noncontrolling interests. The retained panel has 36 sponsors in five sector groups: energy (12), utilities (9), industrials and autos (6), travel (5), and a group of four consumer and technology names (IBM, Coca-Cola, Walmart, Microsoft) intended as low-distress controls. Liabilities are derived as assets minus equity. Filings span 2014\,Q1--2026\,Q1. After requiring four quarters of history for the lagged features and two quarters of forward data for the label, the evaluation sample is \textbf{1{,}418 firm-quarters, 2015\,Q1--2025\,Q3} (energy 454, utilities 333, industrials 244, travel 215, consumer/technology 172).

The universe contains genuine distress: the 2015--16 energy downturn and impairments (Devon, Williams, FirstEnergy), PG\&E's 2019 Chapter~11 filing, the 2020 shock to airlines and cruise lines (Delta, United, Carnival, Royal Caribbean) and energy (Occidental, Apache, Ovintiv), General Electric's 2018 write-downs, and Boeing, whose book equity is thin or negative for almost the whole window.

\subsection{Macro inputs and labels}
Two macro series from FRED~\cite{fred} are joined at annual frequency: the WTI crude price (and its year-on-year change) and the 10-year Treasury yield. Values for 2025--2026 are provisional approximations.

A firm-quarter is in \emph{distress} if (i)~book equity is negative, (ii)~equity has fallen more than $30\%$ below its trailing four-quarter peak, or (iii)~equity has fallen more than $20\%$ quarter-on-quarter. The prediction target is forward-looking: $y_t=1$ if distress occurs in $t{+}1$ or $t{+}2$. The base rate is $11.9\%$ (169 of 1{,}418). These labels are a proxy. They capture balance-sheet erosion that matters for covenant, not default, and they also fire on events that are not credit distress (Section~\ref{sec:limits}).

\section{Methods}
\subsection{Sponsor Vulnerability Index}
Seventeen point-in-time features are computed from the current and lagged filings: leverage, equity ratio, a negative-equity flag, equity drawdown from the trailing peak, quarter- and year-on-year changes in assets and equity, change in leverage, the volatility of the last four quarterly equity changes, WTI level and change, the 10-year yield, and four sector indicators. The classifier is XGBoost~\cite{chen2016} with depth-3 trees, 140 estimators, learning rate $0.06$, row and column subsampling $0.85$, and positive-class weighting equal to the negative/positive ratio.

Evaluation uses \textbf{leave-one-firm-out cross-validation}~\cite{roberts2017}: each of the 36 sponsors is held out in turn, the model is trained on the other 35, and the held-out firm's quarters are scored. All reported metrics are computed on these out-of-fold scores. The operating threshold ($0.72$) is the value that maximizes $F_1$ on the same out-of-fold scores, so threshold-dependent metrics (precision, recall, $F_1$) are mildly optimistic; AUC-ROC and PR-AUC do not depend on it. For the SVI trajectories and caps in Sections~\ref{sec:res-svi}--\ref{sec:res-caps} the model is refit on the full sample; Table~\ref{tab:peaks} compares these in-sample peaks with the out-of-fold ones.

\subsection{Enhanced VaR with a transition-stress overlay}\label{sec:var}
A one-year portfolio return is simulated on $2\times10^5$ paths. The portfolio is a weight $w\in\{0,\,10\%,\,22.5\%\}$ in sponsor securities and $1-w$ in a diversified sleeve of $60\%$ equities ($\mu=6\%$, $\sigma=16\%$) and $40\%$ bonds ($\mu=4\%$, $\sigma=6\%$, $\rho=0.15$). The sponsor return has $\mu=5\%$, correlation $0.45$ with the equity factor, and an assumed sector volatility: $38\%$ (energy), $45\%$ (travel), $30\%$ (industrials), $20\%$ (utilities). The stress overlay, motivated by the NGFS disorderly-transition narrative~\cite{ngfs2023} but \emph{not calibrated to NGFS pathways}, shifts the sponsor drift by $-18$ percentage points for energy, $-10$ for travel and industrials and $-7$ for utilities, and scales sponsor volatility by $1.30$ (energy) or $1.15$ (others). VaR is the negative 5th percentile of the simulated return.

\subsection{Dynamic self-investment cap}
The cap replaces a static base $b=10\%$ (the U.S.\ statutory level~\cite{erisa407}) with
\begin{equation}
\text{cap}_t = b\times m_{\text{ESG}}\times h(\text{SVI}_t),\qquad h(s)=\max(0.1,\;1-0.9\,s).
\end{equation}
In this release $m_{\text{ESG}}$ is a placeholder fixed at $0.8$ for every sponsor, so the effective ceiling is $8\%$ and the floor $0.8\%$.

\section{Results}
\subsection{SVI performance}\label{sec:res-svi}
\begin{figure}[t]\centering\includegraphics[width=\linewidth]{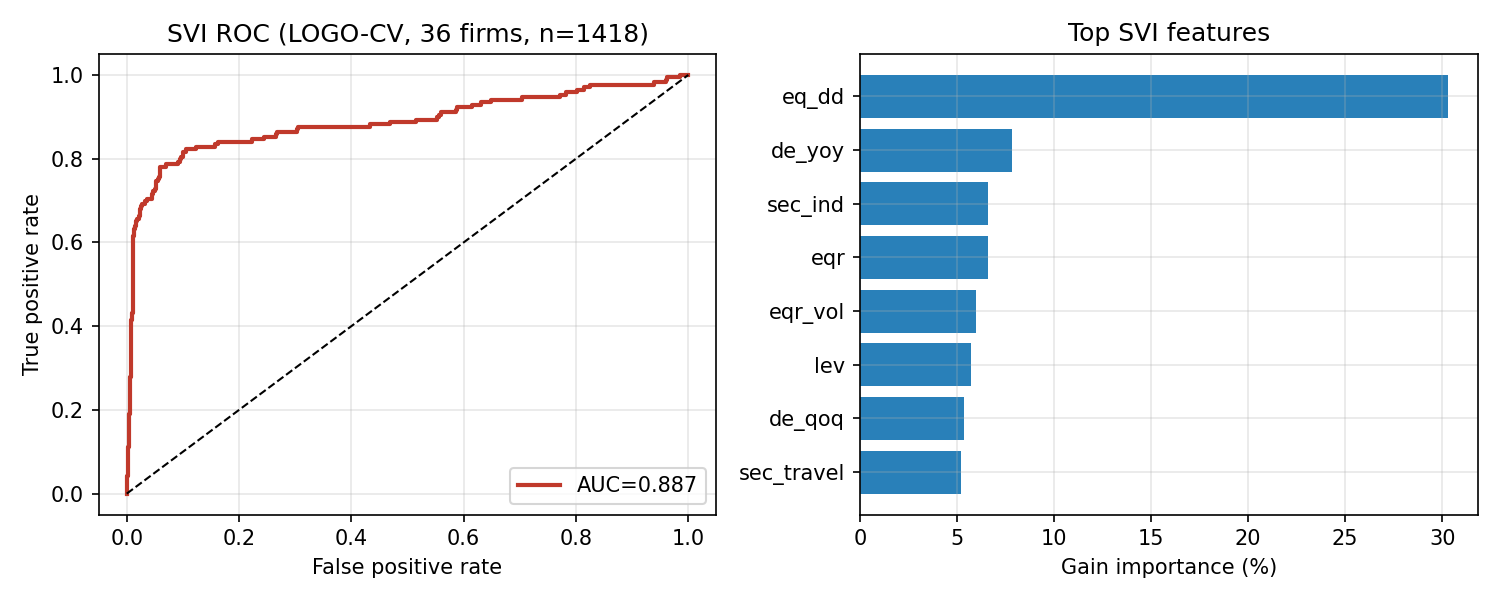}
\caption{Left: out-of-fold ROC curve (leave-one-firm-out, AUC $=0.887$). Right: top features by XGBoost gain.}\label{fig:roc}
\end{figure}

Out of fold, the SVI attains AUC-ROC $0.887$ and PR-AUC $0.711$ against an $11.9\%$ base rate (Table~\ref{tab:svi}, Fig.~\ref{fig:roc}). At the $0.72$ threshold precision is $78.4\%$ and recall $68.6\%$ (confusion matrix: TN 1{,}217, FP 32, FN 53, TP 116); at $80\%$ recall precision falls to $0.53$. The largest contributor by gain is equity drawdown ($30.3\%$), followed by year-on-year equity change ($7.8\%$), the industrials indicator ($6.6\%$), the equity ratio ($6.6\%$), equity-change volatility ($6.0\%$) and leverage ($5.7\%$). The dominance of drawdown is expected, since the label is built partly from it; the model's value lies in anticipating the crossing of the label thresholds before it happens.

\begin{table}[t]\centering\small
\caption{Out-of-fold SVI performance (leave-one-firm-out, 36 sponsors, $n=1{,}418$, 169 positives).}\label{tab:svi}
\begin{tabular}{lcl}
\toprule
Metric & Value & Reading\\\midrule
AUC-ROC & 0.887 & rank-ordering of unseen firms\\
PR-AUC & 0.711 & vs.\ 0.119 base rate\\
Precision @ 0.72 & 78.4\% & about 4 in 5 alerts are followed by distress\\
Recall @ 0.72 & 68.6\% & about 7 in 10 episodes flagged\\
$F_1$ @ 0.72 & 0.732 & threshold chosen to maximize $F_1$\\
Precision @ 80\% recall & 0.53 & cost of high-recall operation\\\bottomrule
\end{tabular}
\end{table}

\begin{figure}[t]\centering\includegraphics[width=\linewidth]{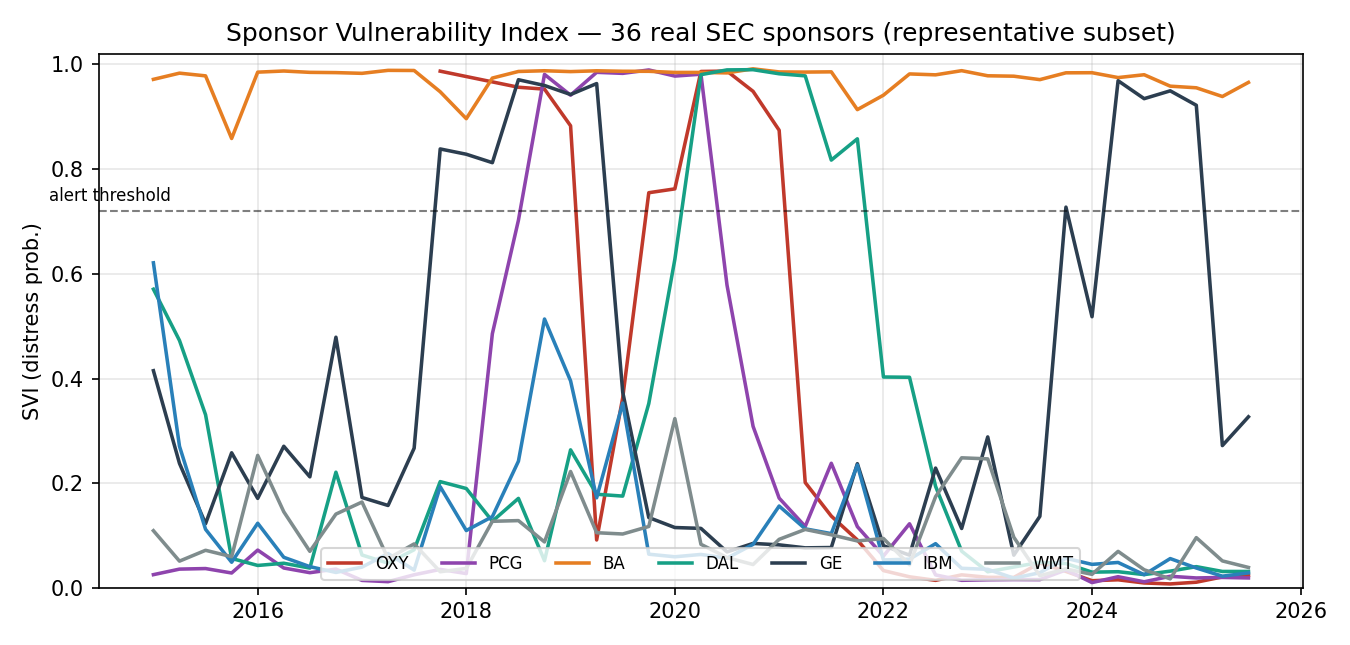}
\caption{SVI (full-sample refit) for a subset of sponsors, 2015--2025. Distress episodes at PG\&E (2018--20), Delta and Occidental (2020--21) and General Electric (2018--19) cross the alert line; Boeing stays above it almost throughout. GE's second excursion (2023--25) coincides with the spin-offs of GE HealthCare and GE Vernova, a book-equity reduction that is not credit distress. IBM and Walmart stay below the line in this fit.}\label{fig:svi}
\end{figure}

\paragraph{Controls and false alarms.} The four control names do not all behave as controls (Table~\ref{tab:peaks}). Walmart and Microsoft stay well below the threshold both in-sample and out of fold. Coca-Cola peaks at $0.76$ (2017\,Q4) and carries two labelled distress quarters, from a book-equity reduction that the label counts as distress although the company was never in credit difficulty. IBM peaks at $0.62$ in-sample but at $0.73$ out of fold (2018\,Q4), crossing the threshold. Among non-control sponsors, Southwest, Dominion, Ford and GM receive out-of-fold scores of $0.77$--$0.85$ in quarters that carry no positive label. These cases account for part of the 32 false positives.

\begin{table}[t]\centering\small
\caption{Peak SVI per sponsor: full-sample refit vs.\ out-of-fold (held-out) scores. Threshold $0.72$.}\label{tab:peaks}
\begin{tabular}{lccccc}
\toprule
Sponsor & Positive qtrs & Peak (refit) & Quarter & Peak (out-of-fold) & Quarter\\\midrule
PG\&E & 8 & 0.99 & 2019\,Q4 & 0.99 & 2020\,Q2\\
Delta & 8 & 0.99 & 2020\,Q4 & 0.99 & 2020\,Q4\\
Boeing & 42 & 0.99 & 2020\,Q4 & 0.96 & 2020\,Q4\\
General Electric & 11 & 0.97 & 2018\,Q3 & 0.98 & 2018\,Q3\\
Coca-Cola & 2 & 0.76 & 2017\,Q4 & 0.79 & 2017\,Q4\\
IBM & 0 & 0.62 & 2015\,Q1 & 0.73 & 2018\,Q4\\
Walmart & 0 & 0.32 & 2020\,Q1 & 0.44 & 2020\,Q1\\
Microsoft & 0 & 0.28 & 2016\,Q1 & 0.39 & 2016\,Q1\\\bottomrule
\end{tabular}
\end{table}

\subsection{Enhanced VaR}
\begin{figure}[t]\centering\includegraphics[width=\linewidth]{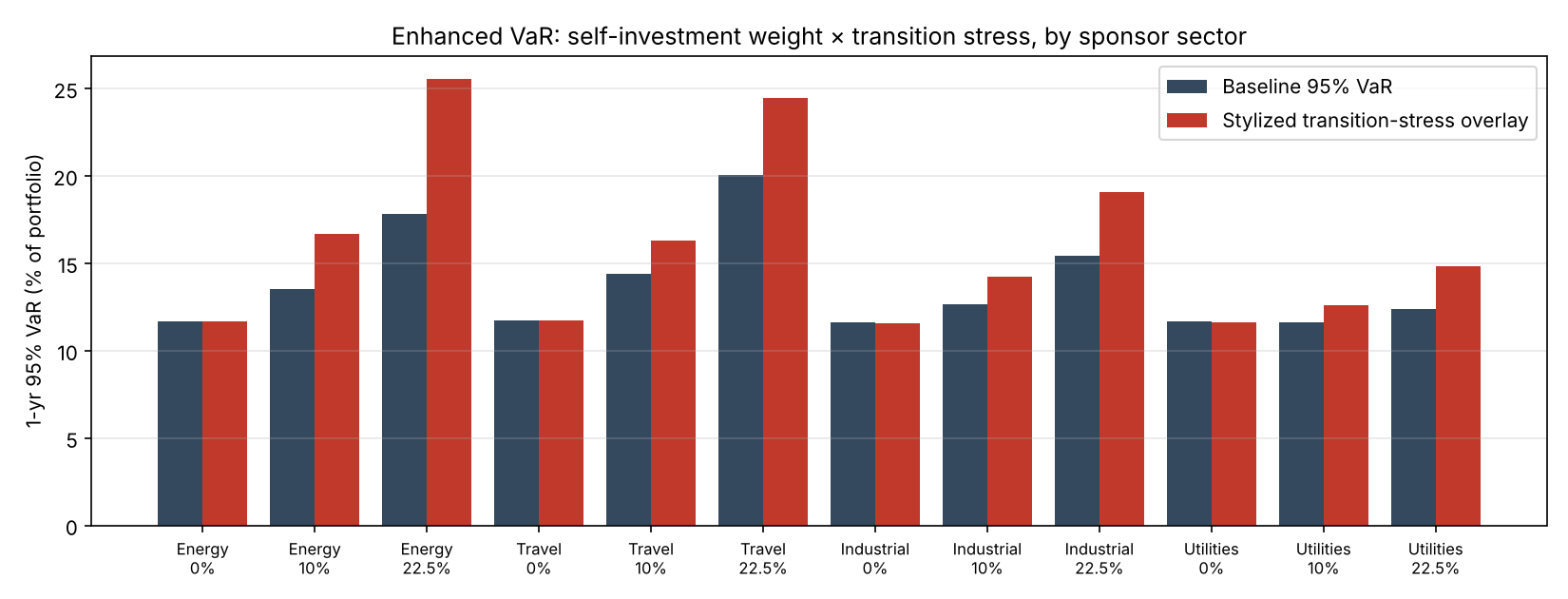}
\caption{Baseline vs.\ stressed one-year 95\% VaR by sponsor sector and self-investment weight.}\label{fig:var}
\end{figure}

Table~\ref{tab:var} and Fig.~\ref{fig:var} report one-year $95\%$ VaR. Without self-investment the stress overlay changes VaR by less than $0.1$ percentage point in every sector, because the diversified sleeve is not stressed. With $10\%$ in sponsor securities the add-on is $+9\%$ (utilities) to $+23\%$ (energy); at $22.5\%$ it is $+20\%$ (utilities), $+22\%$ (travel), $+23\%$ (industrials) and $+43\%$ (energy). Concentration and transition stress therefore have to be assessed jointly: a plan that passes a climate stress test on its diversified holdings can still carry a large climate-sensitive exposure through its sponsor. The magnitudes depend directly on the assumed sector volatilities and stress parameters (Section~\ref{sec:var}). These assumptions differ from realized annualized book-equity volatility in our panel ($50\%$ industrials, $37\%$ travel, $29\%$ energy, $24\%$ utilities, $12\%$ consumer/technology), which we report for reference only.

\begin{table}[t]\centering\small
\caption{One-year 95\% VaR (\% of portfolio): baseline vs.\ stylized transition stress.}\label{tab:var}
\begin{tabular}{lccc}
\toprule
Sponsor sector, self-investment $w$ & Baseline & Stressed & Add-on\\\midrule
Energy, 22.5\% & 17.8 & 25.6 & $+43\%$\\
Energy, 10\% & 13.5 & 16.7 & $+23\%$\\
Industrials, 22.5\% & 15.4 & 19.1 & $+23\%$\\
Travel, 22.5\% & 20.0 & 24.5 & $+22\%$\\
Utilities, 22.5\% & 12.4 & 14.8 & $+20\%$\\
Any sector, 0\% & 11.6--11.7 & 11.6--11.7 & $\approx 0$\\\bottomrule
\end{tabular}
\end{table}

\subsection{Dynamic caps}\label{sec:res-caps}
\begin{figure}[t]\centering\includegraphics[width=\linewidth]{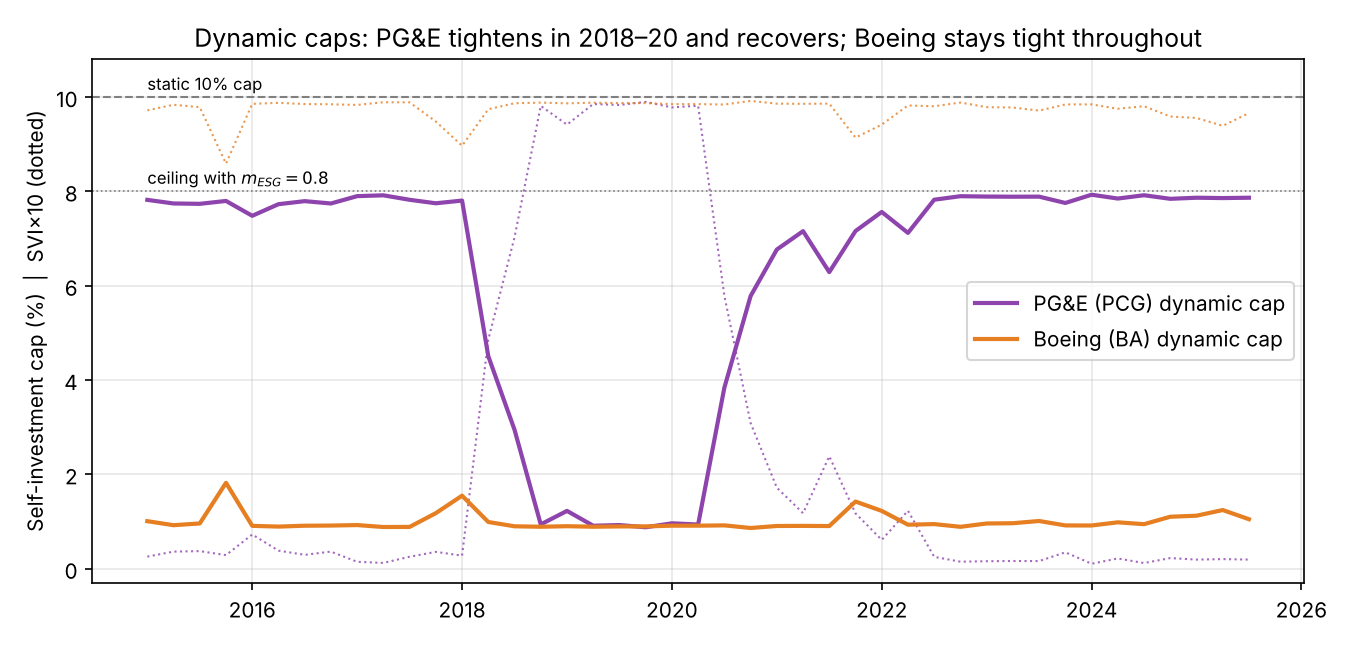}
\caption{Dynamic-cap trajectories (solid) and SVI$\times$10 (dotted) for PG\&E and Boeing, against the static 10\% limit and the 8\% ceiling implied by $m_{\text{ESG}}=0.8$.}\label{fig:caps}
\end{figure}

Table~\ref{tab:caps} and Fig.~\ref{fig:caps} show the cap at each sponsor's peak SVI and in the latest quarter (2025\,Q3). PG\&E's cap falls from about $8\%$ to $0.9\%$ through 2018--20 and returns to $7.9\%$ after emergence from bankruptcy; Delta follows the same path around 2020. Boeing's cap stays near $1\%$ for the entire sample, reflecting persistently thin or negative book equity. For Boeing the SVI is therefore a standing warning rather than an early signal. General Electric's cap is $5.6\%$ in 2025\,Q3 after the spin-off episode. Moderately elevated sponsors are tightened in proportion: Exxon's peak SVI of $0.46$ in 2020\,Q4 implies a $4.7\%$ cap, Walmart's $0.32$ implies $5.7\%$.

\begin{table}[t]\centering\small
\caption{Dynamic cap (base 10\%, $m_{\text{ESG}}=0.8$) at peak SVI vs.\ latest quarter (2025\,Q3).}\label{tab:caps}
\begin{tabular}{lccccc}
\toprule
Sponsor & Peak quarter & Peak SVI & Cap at peak & SVI 2025\,Q3 & Cap 2025\,Q3\\\midrule
PG\&E & 2019\,Q4 & 0.99 & 0.9\% & 0.02 & 7.9\%\\
Delta & 2020\,Q4 & 0.99 & 0.9\% & 0.03 & 7.8\%\\
Boeing & 2020\,Q4 & 0.99 & 0.9\% & 0.97 & 1.0\%\\
General Electric & 2018\,Q3 & 0.97 & 1.0\% & 0.33 & 5.6\%\\
Exxon Mobil & 2020\,Q4 & 0.46 & 4.7\% & 0.06 & 7.5\%\\
Walmart & 2020\,Q1 & 0.32 & 5.7\% & 0.04 & 7.7\%\\\bottomrule
\end{tabular}
\end{table}

\section{Discussion and limitations}\label{sec:limits}
Three conclusions follow. Sponsor distress, in the book-equity sense relevant to covenant, is predictable from ordinary filings with one to two quarters of lead time, and the signal generalizes to firms the model has not seen. Self-investment and transition stress interact and should be tested together. Risk limits can be made responsive rather than static at little cost in transparency, since the cap is a closed-form function of a published score.

The limitations matter as much as the results:
\begin{itemize}
\item \textbf{Labels.} Book-equity thresholds proxy for credit events; they flag non-credit reductions in book equity (Coca-Cola 2017) and spin-offs (GE 2023--25) as distress. Market-based and credit-event labels (rating downgrades, CDS spreads, defaults) are the most important extension.
\item \textbf{Concentrated positives.} Boeing contributes 42 of the 169 positives. Leave-one-firm-out validation prevents Boeing's quarters from informing its own score, but one firm still shapes what the model learns about the other 35.
\item \textbf{Threshold and refit.} The $0.72$ threshold is tuned on the evaluation scores, and the trajectories and caps use a full-sample refit; a deployed system should fix the threshold on a separate period and use strictly out-of-sample scores.
\item \textbf{Macro inputs.} WTI and the 10-year yield enter as annual averages; 2025--26 values are provisional.
\item \textbf{VaR parameters.} Sponsor volatilities and the stress overlay are assumptions, not estimates; the ESG multiplier is a constant placeholder.
\item \textbf{Scope.} The panel is 36 U.S.\ GAAP filers. Extending it to the S\&P~500, to Form~5500 plan data~\cite{dol5500}, and to non-U.S.\ reporting is left for future work.
\end{itemize}

\section{Conclusion}
We presented a reproducible, public-data early-warning pipeline for pension-sponsor distress. On 1{,}418 firm-quarters of SEC filings the SVI reaches an out-of-fold AUC-ROC of $0.887$; a stress-augmented VaR shows concentration and transition stress interacting across sectors; and a dynamic cap converts the score into a responsive limit. The same method underlies \emph{Faro}, a Spanish- and Portuguese-language demonstrator for Latin American pension supervision built on illustrative data (\url{https://digitaleconomylab.org/faro}).

\paragraph{Code and data availability.} Code, extracted EDGAR facts, the panel, results and figures are released at \url{https://github.com/rav11l/pension-research-SponsorVulnerabilityModel} and archived on Zenodo under the concept DOI \href{https://doi.org/10.5281/zenodo.21394088}{10.5281/zenodo.21394088}, which resolves to the latest version.

\end{document}